\documentclass[a4paper,11pt]{article}
\usepackage{jheppub} 
\usepackage{lineno}
\usepackage{xspace}
\usepackage{siunitx}

\newcommand{\pt}{\mbox{$p_{\rm{T}}$}\xspace}
\newcommand{\snn}{\mbox{$\sqrt{s_{\rm NN}}$}\xspace}
\newcommand{\mpt}{\mbox{$\langle p_{\rm T}\rangle$}\xspace}
\newcommand{\npart}{\mbox{$\rm \langle N_{part}\rangle$}\xspace}
\newcommand{\apx}{\mbox{$\approx$}\xspace}

\title{\boldmath Strangeness Production in $\sqrt{s_{\rm NN}}=3$ GeV Au+Au Collisions at RHIC}

\collaboration{\includegraphics[height=17mm]{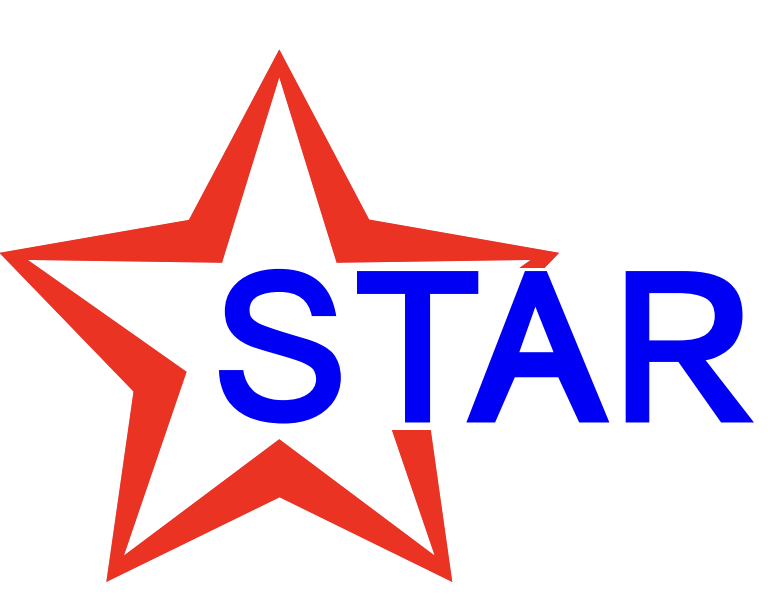}\\[6pt]
 The STAR Collaboration}

\emailAdd{star-publication@bnl.gov}

\abstract{We report multi-differential measurements of strange hadron production ranging from mid- to target-rapidity in Au+Au collisions at a center-of-momentum energy per nucleon pair of $\sqrt{s_{\rm NN}}=3$ GeV with the STAR experiment at RHIC. $K^0_S$ meson and $\Lambda$ hyperon yields are measured via their weak decay channels. Collision centrality and rapidity dependences of the transverse momentum spectra and particle ratios are presented. Particle mass and centrality dependence of the average transverse momenta of $\Lambda$ and $K^0_S$ are compared with other strange particles, providing evidence of the development of hadronic rescattering in such collisions. The 4$\pi$ yields of each of these strange hadrons show a consistent centrality dependence. Discussions on radial flow, the strange hadron production mechanism, and properties of the medium created in such collisions are presented together with results from hadronic transport and thermal model calculations.}

\begin{document}
\maketitle
\flushbottom

\section{Introduction}
\label{sec:intro}

The quantum chromodynamics (QCD) phase diagram~\cite{Andronic:2017pug,Fukushima:2020yzx}, often represented in terms of temperature ($T$) vs. baryonic chemical potential ($\mu_{\rm B}$), is characterized by the phenomenologically determined boundary $T=T_{\rm ch}(\mu_{\rm B})$, where $T_{\rm ch}$ refers to the chemical freeze-out temperature. A highly dense medium, called the Quark-Gluon Plasma (QGP), is expected to be created at sufficiently high temperatures such as those produced at the Relativistic Heavy Ion Collider (RHIC) at its top energy and at the energies of the Large Hadron Collider (LHC), where high energy density and vanishing $\mu_{\rm B}$ are achieved. According to Lattice QCD calculations~\cite{Aoki:2006we} and recent experimental results~\cite{STAR:2005gfr,STAR:2021rls}, the transition from QGP to hadronic matter is a smooth crossover with a pseudo-critical temperature at $\mu_{\rm B} = 0$ of $T_{\rm c} = 156.5 \pm 1.5$ MeV~\cite{HotQCD:2018pds}. In low energy collisions where the net-baryon density is large, a first-order phase transition between QGP and hadronic matter has been predicted~\cite{Asakawa:1989bq}. The first-order phase transition line is expected to end at a critical point where a smooth crossover begins~\cite{Fukushima:2010bq}. Recent discussions on chemical freeze-out can be found in Ref.~\cite{Muller:2022qsm}.

Strangeness enhancement was proposed as a signature for the formation of the QGP in high-energy nuclear collisions where thermalized strange quarks are created during QGP evolution and coalesce into color-singlet hadrons during hadronization\cite{Koch:1986ud}. Such enhancement, first measured in Pb-Pb collisions at SPS~\cite{WA97:1999uwz} has been observed experimentally over a wide range of collision energies and system sizes. The system reaches the grand canonical limit in central heavy-ion collisions at high energies, as shown by results from RHIC and LHC~\cite{STAR:2008med, STAR:2017sal, Andronic:2017pug,ALICE:2016fzo,ALICE:2020nkc}. The sixth order net-proton cumulant ratios~\cite{STAR:2021rls} from central Au+Au collisions at a center-of-momentum energy per nucleon pair of $\sqrt{s_{\rm NN}}=200$ GeV are consistent with Lattice QCD calculations for the formation of the QGP~\cite{Bazavov:2020bjn,Borsanyi:2018grb}. In the high net-baryon density region ($\mu_{\rm B}>$ 500 MeV), strangeness production serves as a crucial probe for analyzing the hot and dense nuclear matter created in heavy-ion collisions and for studying the nuclear equation of state (EoS)~\cite{KAOS:2000ekm}. At lower collision energies, local strangeness conservation needs to be enforced for statistical hadronization models in order to describe the recent data on $\phi$ and $\Xi^{-}$ production in $\sqrt{s_{\rm NN}}=3$ GeV Au+Au collisions~\cite{STAR:2021hyx}. Local canonical equilibrium is a characteristic of the hadronic system where the correlation length for strangeness production is significantly smaller than the radius of the fireball~\cite{Kraus:2007hf}. Results on collective flow and high-order proton cumulant ratios from $\sqrt{s_{\rm NN}}=3$ GeV collisions also support the hadron-dominant nature of the produced medium~\cite{STAR:2021yiu,STAR:2021ozh,STAR:2021fge}.
 
In this paper, we report strange hadron ($\Lambda$ and $K_S^0$) production in Au+Au collisions at $\sqrt{s_{\rm NN}}=3$ GeV with the STAR experiment at RHIC.  Transverse momentum ($p_{\rm T}$) and rapidity ($y$) distributions of $\Lambda$ and $K_S^0$ are presented. The paper is organized as follows: Section~\ref{sec:expt} describes the experimental setup, data sets, analysis details including signal extraction and efficiency corrections. Systematic uncertainties are discussed in Section~\ref{sec:systematic}. The strange hadron transverse momentum spectra, rapidity distributions and mean transverse momentum are presented for different centrality intervals in Section~\ref{sec:result}. Various particle ratios along with thermal and transport model comparisons are presented in the same section. Finally, we summarize our findings in Section~\ref{sec:summary}. 
\section{Experimental setup and data analysis}
\label{sec:expt}

The dataset used in this analysis was collected using the fixed-target (FXT) setup~\cite{Meehan:2016iyt, STAR:2021beb, STAR:2020dav} at the Solenoidal Tracker At RHIC (STAR) experiment during the 2018 RHIC run. A single beam was provided by RHIC with a total energy equal to 3.85 GeV/nucleon. The gold target, of thickness 0.25 mm corresponding to a 1\% interaction probability, is installed inside the vacuum pipe, 2 cm below the center of the beam axis, and located \SI{200.7}{cm} west of the STAR detector’s center. The main detector used in this analysis is the Time Projection Chamber (TPC), which has full azimuthal coverage within a pseudorapidity range of $-2<\eta<0$ in FXT mode~\cite{Anderson:2003ur}. In addition to its track reconstruction and momentum determination capabilities, the TPC provides particle identification for charged particles by measuring their ionization energy loss per unit length ($dE/dx$) in the TPC gas. Details and performance of the energy loss and particle identification method are explained in Ref.~\cite{Shao:2005iu}. The offline reconstructed primary vertex position is required to be within 2 cm of the target  along the beam direction and within a radius of 1.5 cm in the transverse plane from the center of the target in order to eliminate possible backgrounds arising from beam interactions with the vacuum pipe. Approximately $2.6\times10^8$ minimum bias (MB) events pass the selection criteria and are used in this analysis. The centrality of the collision is determined using the number of reconstructed charged-particle tracks in the TPC acceptance in conjunction with a Monte Carlo Glauber model simulation~\cite{Ray:2007av}. 

$\Lambda$ and $K^0_S$ hadrons are reconstructed via their weak decay channels $\Lambda \rightarrow p + \pi^-$ (branching ratio B.R. $= 63.9\%$) and $K^0_S \rightarrow \pi^ + \pi^-$ (B.R. $= 69.20\%$)~\cite{ParticleDataGroup:2020ssz}, respectively. We require the reconstructed tracks to have at least 15 measured space points in the TPC (out of a maximum possible 45) and a minimum reconstructed transverse momentum of 100 $\rm{MeV}/c$ to ensure good track quality. To suppress split tracks, the ratio of the number of hits on a track to the maximum possible number of hits that this track may possess must be larger than 0.52. Particle identification for $\pi^-$, $\pi^+$ and proton is achieved by measuring the $dE/dx$ in the TPC. In this analysis, a cut of $|n\sigma|<3$ is used in particle identification. Here $n\sigma$ is defined as
\begin{linenomath*}
\begin{equation*}
n\sigma=\frac{1}{\sigma_{R}}\log\frac{\langle dE/dx\rangle}{\langle dE/dx\rangle_{\rm Bichsel}},
\end{equation*}
\end{linenomath*}
where $\langle dE/dx\rangle_{\rm Bichsel}$ is the expected $\langle dE/dx\rangle$ from the Bichsel function~\cite{Bichsel:2006cs} for that particle species at a given momentum and $\sigma_{R}$ is the resolution of $\log(\langle dE/dx\rangle / \langle dE/dx\rangle_{\rm Bichsel})$ of the TPC\cite{Shao:2005iu}. The KFParticle package~\cite{kfptc}, a particle reconstruction software package based on the Kalman Filter, is used for the reconstruction of $\Lambda$ and $K^0_S$. Various topological variables, such as the distances of closest approach (DCA) between the mother/daughter particles to the primary vertex, and the DCA between the two daughters~\cite{kfptc2}, are examined. Cuts on these topological variables are applied to the signal candidates to optimize the statistical significance. 

\begin{figure}[htb]
\centering
\centerline{\includegraphics[width=0.8\textwidth]{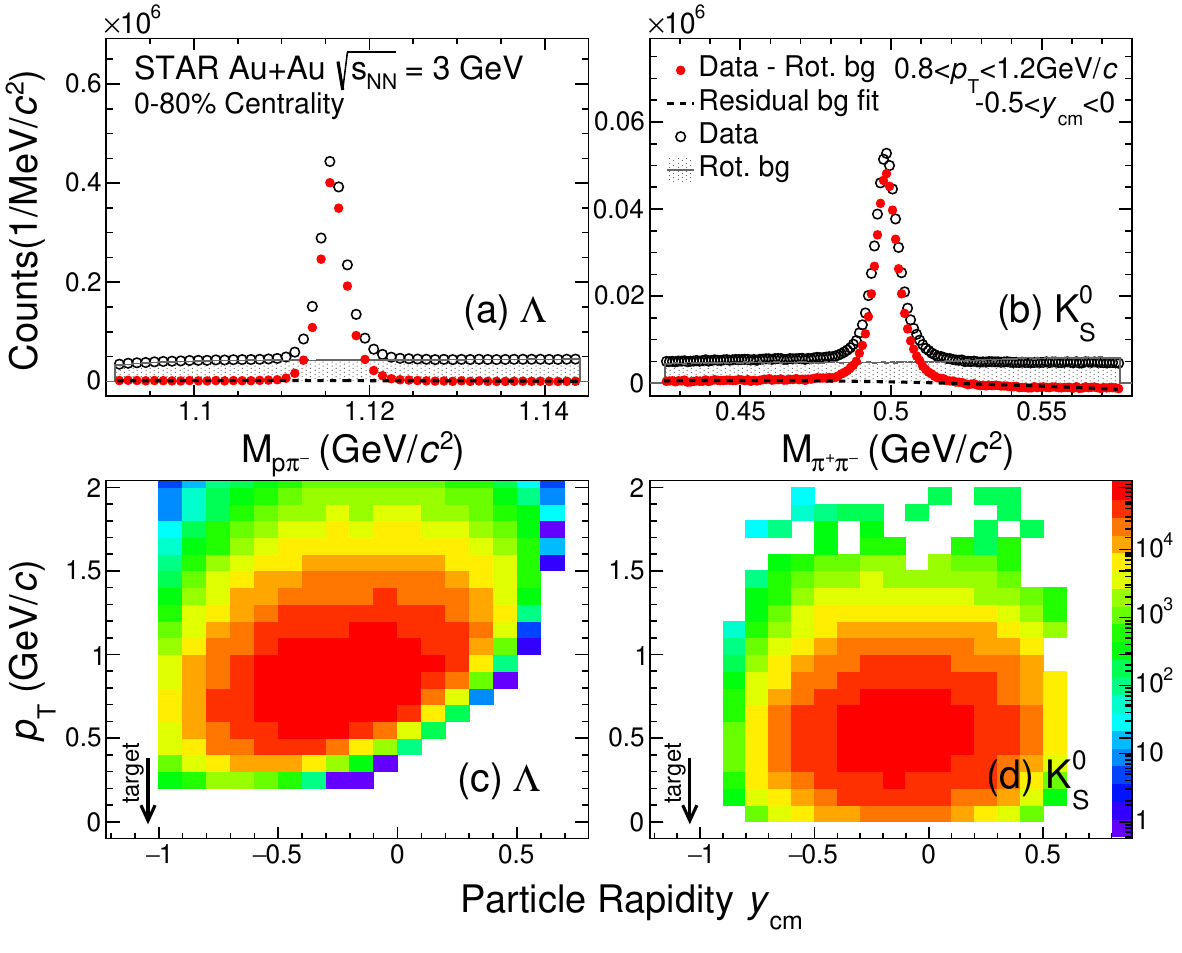}}
\caption{Invariant mass distributions (black open circles) of (a) $p\pi^-$ and (b) $\pi^+\pi^-$ in Au+Au collisions at $\sqrt{s_{\rm NN}}$= 3 GeV. The grey shaded histogram represents the rotated background distribution used to estimate the combinatorial background. The red solid circles depict the $\Lambda$ (a) and $K^0_S$ (b) signals after subtracting the combinatorial background. The black dashed line represents the polynomial function which is an estimate for the residual background. Reconstructed $\Lambda$ (c) and $K^0_S$ (d) acceptances are shown as $\it{p}_{\rm T}$ vs. rapidity in the center-of-momentum frame ($y_{\rm cm}$) . The arrow indicates the target rapidity.}
\label{pid}
\end{figure}

Figure~\ref{pid} shows the invariant mass distributions of (a) $p\pi^-$ pairs and (b) $\pi^+\pi^-$ pairs in the $p_{\rm T}$ region (0.8 – 1.2) $\rm{GeV}/c$ in 0-80\% collisions. The combinatorial background is estimated using a rotation technique, in which all $\pi^-$ tracks in a single event are rotated by fixed angles ($\pi/2$, $\pi$, and $3\pi/2$) in the transverse plane. The invariant mass distributions obtained from this rotation technique are then scaled to match the number of $p\pi^-$ pairs or $\pi^+\pi^-$ pairs in the off-peak regions ($1.13-1.15$ GeV/$c^{2}$ for $p\pi^-$ pairs, $0.45-0.47$, and $0.53-0.55$ GeV/$c^{2}$ for $\pi^+\pi^-$ pairs). The background shape is well reproduced using the rotation technique for both $\Lambda$ and $K^0_S$ as shown in Fig.~\ref{pid} (a) and (b). The combinatorial background is subsequently subtracted from the data in 2D phase space ($p_{\rm T}$ and rapidity $y_{\rm cm}$) in the collision center-of-momentum frame. After subtracting the combinatorial background, the resulting distributions, shown as red solid circles, are fitted with double-Gaussian (a sum of two Gaussian functions with the same mean value) plus a polynomial function to determine the signal peak width as well as the shape of the remaining residual correlated background. These residual backgrounds originate from unavoidable particle mis-identification. For example, a proton from a $\Lambda$ decay misidentified as a $\pi^+$ could be combined with the $\pi^-$ daughter, thus contributing to the residual background in $K^0_S$ reconstruction. In order to suppress such residual background, veto cuts are introduced. For $K^0_S$ reconstruction, the $\pi^+$ daughter of a $K^0_S$ candidate is assumed to be a $p$ and the invariant mass of the pair is recalculated.  If it falls inside the invariant mass peak of $\Lambda$, then the $K^0_S$ candidate is rejected. The $\Lambda$ and $K^0_S$ raw yields are obtained via histogram bin counting from the invariant mass distributions with all backgrounds subtracted within mass windows of width 3$\sigma$ from the mean ($\mu$), where the $\mu$ and $\sigma$ of the double Gaussian are obtained from the fit explained previously. The $\Lambda$ and $K^0_S$ acceptances represented as $p_{\rm T}$ versus rapidity in the center-of-momentum frame are shown in Fig.~\ref{pid} (c) and (d), respectively. The target is located at $y_{\rm cm}$ = -1.05, using the convention where the beam travels in the positive direction.

The raw yields of each particle are obtained in $p_{\rm T}$ and rapidity bins for different centrality selections, and are subsequently corrected for acceptance and efficiency. The TPC acceptance and tracking efficiency corrections account for tracks not falling inside the TPC acceptance or failing to meet the single track selection criteria, while the topology cut efficiency correction accounts for track pairs failing to meet the decay topology criteria. These corrections are estimated using a Monte Carlo (MC) simulation. The simulated particles are propagated through a TPC detector response simulator using GEANT3~\cite{brun1987geant} to produce simulated track data. These simulated tracks are embedded into real events, and reconstructed using the same analysis chain as for real data.

The $p_{\rm T}$ spectra of each strange hadron are obtained by dividing the event-normalized raw yield in a certain $p_{\rm T}$ interval by the corresponding acceptance and reconstruction efficiencies. For the $\Lambda$ $p_{\rm T}$ spectra, weak decay feed-down contributions from $\Xi^{-}$ and $\Xi^{0}$ need to be considered. The feed-down contribution from $\Omega$ is found to be negligible and is neglected in this study. Following the procedure from Ref.~\cite{STAR:2019bjj}, the feed-down contributions from such decays are estimated with the help of embedding data. The MC $\Xi^{-}$ and $\Xi^{0}$ yields are weighted with realistic kinematic distributions: the differential yield of $\Xi^{-}$ is taken from Ref.~\cite{STAR:2021hyx} while the $\Xi^{0}$ is assumed to have the same $p_{\rm T}$, rapidity, and centrality dependence as $\Xi^{-}$ with a ratio of $\Xi^{0}/\Xi^{-}=0.9$ estimated from thermal model THERMUS~\cite{wheaton2009thermus}, which will be described in detail in the next section. The decayed $\Lambda$s  from MC $\Xi^{-}$ and $\Xi^{0}$ are then reconstructed using the same reconstruction chain as used for real data analysis. The feed-down contributions are determined as a function of $p_{\rm T}$, rapidity and centrality. They decrease with increasing $p_{\rm T}$ and from mid-rapidity to backward rapidity, and are found to be small ($<4\%$ in 0-10\% collisions and $<1\%$ in 60-80\% collisions). These feed-down contributions are subtracted from the inclusive $\Lambda$ yield as a function of $p_{\rm T}$, rapidity, and centrality. It should be noted that $\Sigma^{0}$ baryons decay to $\Lambda$ via $\gamma$ emission with a $100\%$ branching ratio. Due to the $\Sigma^{0}$'s short lifetime, the $\Lambda$s arising from its decays cannot be distinguished from primordial $\Lambda$s. In this work, we do not subtract feed-down contributions arising from $\Sigma^{0}$ decays.

Due to limited detector acceptance at low $p_{\rm T}$, and reduced statistics at high $p_{\rm T}$, the spectra cannot be measured in these regions and extrapolation is needed in order to obtain the $p_{\rm T}$-integrated yield ($dN/dy$) as well as the mean transverse momentum ($\langle p_{\rm T}\rangle$). The blast-wave model~\cite{Schnedermann:1993ws} is used for fitting $p_{\rm T}$ spectra in the measured region and extrapolating them to the unmeasured regions. The $p_{\rm T}$ spectra of strange hadrons produced in Au+Au collisions at the RHIC beam energy scan (BES) energies are well described by the blast-wave model~\cite{STAR:2019bjj}. This model assumes that particles decouple from a system in local thermal equilibrium with temperature $T$, that expands both longitudinally and transversely. The longitudinal expansion is taken to be boost-invariant and the transverse expansion is defined in terms of a transverse flow velocity profile. The transverse velocity profile can be parameterized according to a power law: $\beta_{T}(r) = \beta_{S}(r/R)^n$ where $\beta_{S}$ is the maximum surface flow velocity and the exponent $n$ describes the evolution of the flow velocity (flow profile) from any radius $r$ up to $R$ $(r<R)$, where $R$ is the maximum radius of the expanding source at thermal freeze-out. In this analysis, a linear $(n = 1)$ $r$-dependence of the transverse flow velocity is used. The extrapolated region contributions vary from 30\% to 60\% for $\Lambda$ and 5\% to 30\% for $K^0_S$ of the $p_{T}$-integrated yields. The final $dN/dy$ is obtained by summing the data in the measured region and the integral of the fitted function in the unmeasured region. 
The $\langle p_{\rm T}\rangle$ is determined as follows:
\begin{linenomath*}
\begin{equation*}
\langle p_{\rm T}\rangle = \frac{\int \pt\frac{dN}{d\pt}d\pt}{\int \frac{dN}{d\pt}d\pt}.
\end{equation*}
\end{linenomath*}
The same functional forms used to determine the total yield are used for the calculation of $\langle p_{\rm T}\rangle$. Total hadron yields are obtained by integrating the rapidity distribution in the measured region and using a three-Gaussian~\cite{NA49:2016qvu} fit for extrapolating to the unmeasured region, where one Gaussian has a mean at $y_{\rm cm}=0$ and the other two have means symmetric about $y_{\rm cm}=0$ with the same amplitudes and widths.

\section{Systematic uncertainties}
\label{sec:systematic}
We consider four major sources of systematic uncertainty in the $p_{\rm T}$ spectra: imperfect description of topological variables in the simulations, the track selection, the signal extraction technique, and uncertainty in the global tracking efficiency. The first three contributions are estimated by varying the topological cuts, the TPC track quality selection cuts (minimum number of TPC hits), and the background subtraction method. The systematic uncertainty due to global tracking efficiency has been estimated assuming 5\% uncertainty in the single charged-particle tracking efficiency~\cite{star2023production}, resulting in 10\% systematic uncertainty for $\Lambda$ and $K^0_S$ with two decay daughter 
particles. These four uncertainties are assumed to be uncorrelated with each other and are added in quadrature. For $\Lambda$, based on the difference between THERMUS and ART\cite{PhysRevC.106.024902} calculations, we vary the estimated $\Xi^{0}$ yield by $\pm 30\%$ to estimate the uncertainty from the feed-down correction, which is less than $1\%$ in all considered kinematic regions. The systematic uncertainties of different sources are listed in Table~\ref{tab:sysError} for measurements in 0-10\% centrality. The systematic uncertainties are similar for different centralities. For $dN/dy$ and $\langle p_{\rm T}\rangle$ measurements, the systematic uncertainty in the extrapolation to the unmeasured region needs to be considered. Different functional forms, such as the $m_{\rm T}$-exponential function and the Levy function, are also used for extrapolation. The variations compared to that from the default function are assigned as the systematic uncertainty. The systematic uncertainty of the $dN/dy$ extrapolation to obtain the yields for the full phase space (4$\pi$ yields) is estimated based on the difference between fits to the data using a sum of three-Gaussian functions and the $dN/dy$ shape from the Ultra-relativistic Quantum Molecular Dynamics model (UrQMD)~\cite{Bleicher:1999xi}, that is described in detail in the next section.

\begin{table}[htbp]
\centering
 \caption{Summary of systematic uncertainties for the $\Lambda$ and $K_S^{0}$ $dN/dy$ measurements in 0-10\% Au+Au collisions at $\sqrt{s_{\rm{NN}}}$ = 3.0\,GeV. The ranges indicate the variation of the systematic uncertainty among rapidity bins.}
\begin{tabular}{c|cc}
    \hline
    \textbf{Source} & $\Lambda$ & $K_S^0$ \\
    \hline
    Topological cuts & $ 0.7-3.4\%$ & $1.1-3.1\%$ \\
    Track selection & $0.1-0.5\%$ & $0.6-4.6\%$ \\
    Tracking efficiency & $10\%$ & $10\%$ \\
    Signal extraction & $0.4-0.8\%$ & $0.1-0.7\%$ \\
    Extrapolation & $3.6-11\%$ &  $0.2-1.6\%$ \\
    Feed-down correction & $0.4-0.8\%$ &  N/A \\ \hline 
    \textbf{Total} & $10.8-15.3\%$ &  $10.2-11.6\%$ \\ \hline
\end{tabular}
\label{tab:sysError}
\end{table}

\section{Results and discussion}
\label{sec:result}
\begin{figure}[htbp]
\centering
\centerline{\includegraphics[width=0.9\textwidth]{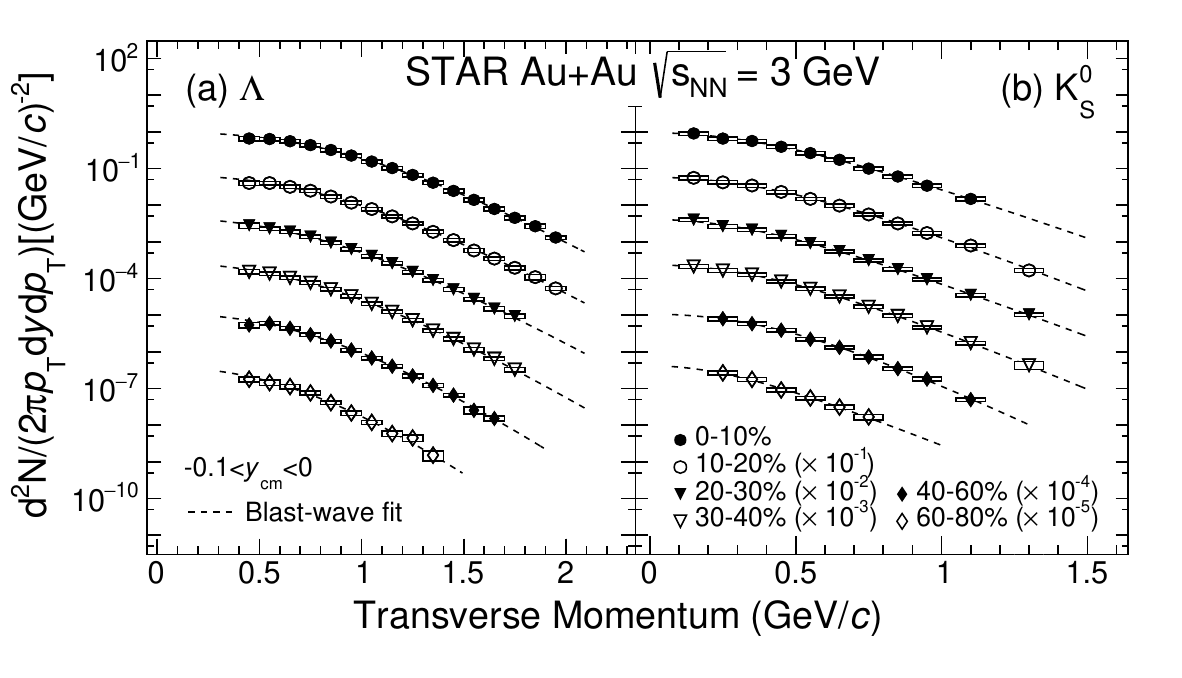}}
\caption{Transverse-momentum spectra of $\Lambda$ (a) and $K^0_S$ (b) at mid-rapidity from different centrality bins in Au + Au collisions at $\sqrt{s_{\rm NN}}$ = 3 GeV. The spectra are scaled by consecutive factors of $10^{-1}$ for each centrality bin as indicated in the legend. The vertical lines and boxes represent the statistical and systematic uncertainties, respectively. The dashed curves represent fits to the data using the blast-wave model.}
\label{dndptdy}
\end{figure}

Figure~\ref{dndptdy} shows the acceptance and efficiency corrected $\Lambda$ (a) and $K^0_S$ (b) invariant yields as a function of $p_{\rm T}$ for $-0.1<{\it y}_{\rm cm}<0$ in six centrality intervals from $\sqrt{s_{\rm NN}}$ = 3 GeV Au$+$Au collisions. Dashed lines depict fits to the spectra with the blast-wave function. For the $\Lambda$ and $K^0_S$ invariant yields in other rapidity regions, please refer to Sec.~\ref{sec:suppl}.

\begin{figure}[htbp]
\centering
\centerline{\includegraphics[width=0.9\textwidth]{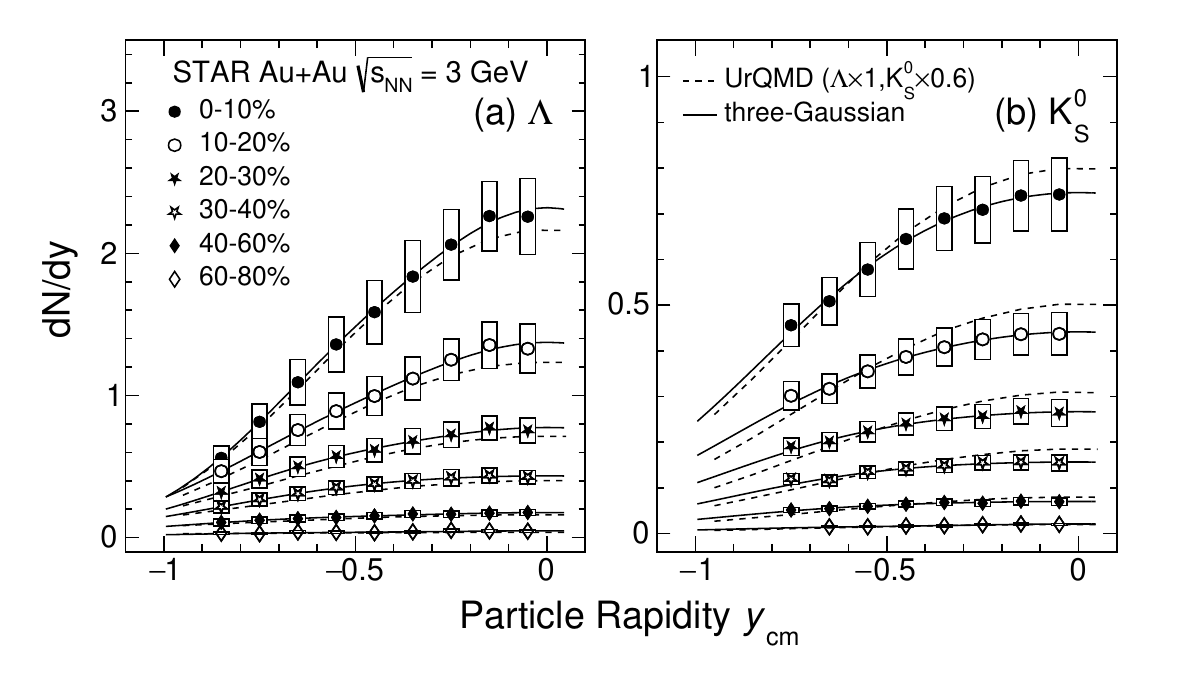}}
\caption{The rapidity dependence of $dN/d\it{y}$ of particles for different centrality bins in Au+Au collisions at $\sqrt{s_{\rm NN}}$ = 3 GeV. The vertical lines and boxes represent the statistical and systematic uncertainties, respectively. The solid lines represent the three-Gaussian function that fit the data points. The dashed lines are the calculations from hadronic transport model UrQMD~\cite{Bleicher:1999xi}.}
\label{dndy}
\end{figure}
The $p_{\rm T}$-integrated rapidity distributions $dN/dy$ are displayed in Fig.~\ref{dndy} for Au+Au collisions at $\sqrt{s_{\rm NN}}$ = 3 GeV for different centralities. The solid lines represent fits using the three-Gaussian function. They are used to extrapolate to the unmeasured rapidity region for calculating 4$\pi$ yields where the positive rapidity region is a reflection of negative region. 

The UrQMD model is a microscopic hadronic transport model based on the propagation and 2-body scattering of hadrons. In order to compare the shape of the rapidity distributions, the model curves for $K^0_S$ are scaled by a factor of 0.6. The UrQMD calculation gives a fair description of the rapidity dependence and the centrality dependence of both $\Lambda$ and $K^0_S$, although it overestimates the absolute yield of $K^0_S$. Comparing the $K^0_S$ rapidity distribution to that of the $K^{-}$~\cite{STAR:2021hyx}, it is found that the width of the distributions is narrower for $K^{-}$ compared to $K^0_S$ for all centralities.  In a hadronic medium, $K^-$ is produced from pair production $NN\rightarrow NNK^{-}K^{+}$, which requires a total energy threshold $\sqrt{s_{\rm {NN,threshold}}}=E_s=2.86$ GeV, while $K^0$ has contributions from both pair production and associated production $NN\rightarrow N\Lambda K^{0}$, with a nucleon-nucleon ($NN$) production threshold of $E_s = 2.56$ GeV which is lower than that of pair production. The different thresholds and production kinematics can lead to a wider rapidity width for $K_{S}^0$ compared to $K^{-}$\cite{E-802:1998,KaoS:1997}.

\begin{figure}[htbp]
\centering
\centerline{\includegraphics[width=0.6\textwidth]{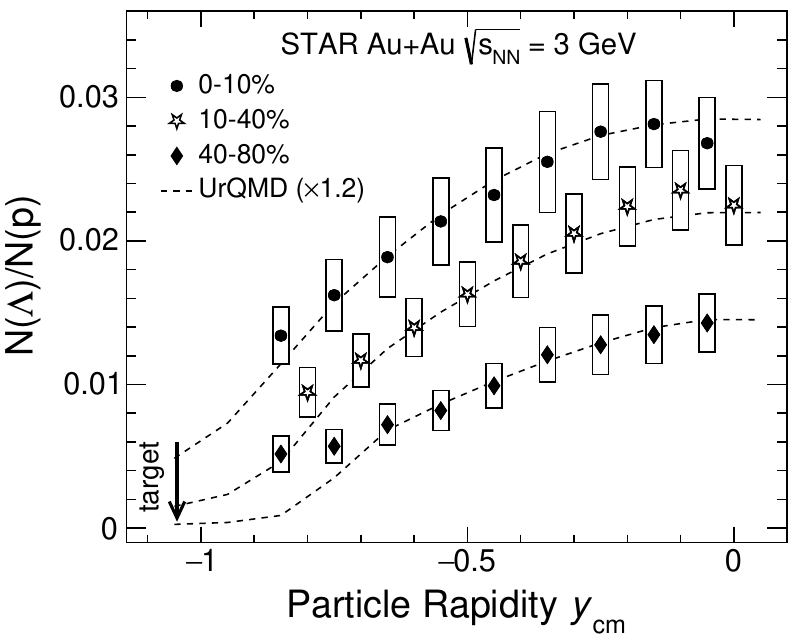}}
\caption{Rapidity dependence of $\Lambda/p$ for different centrality bins in Au+Au collisions at $\sqrt{s_{\rm NN}}$ = 3 GeV. Vertical lines represent statistical uncertainties, while boxes represent systematic uncertainties. The 10-40\% centrality data points are shifted to the right for better visibility. The curves represent the calculations from UrQMD and are scaled up by 20\% to match the data at mid-rapidity. Proton data are taken from Ref.~\cite{star2023production}.}
\label{ldvsp}
\end{figure}

The comparison of the kinematic distributions of the similar mass baryons, $p$~\cite{star2023production} and $\Lambda$, may help us gain more information about the production of strangeness. Also, the $\Lambda/p$ ratio is a necessary input to compute the strangeness population factor, $S_3={}^{3}_{\Lambda}\rm{H}/({}^3He\times \Lambda/\it{p})$, which can give insight into the hypernuclei production mechanisms~\cite{PhysRevC.107.014912}. Figure~\ref{ldvsp} shows the rapidity dependence of the $\Lambda/p$ ratio in different centrality selections. The data are compared to UrQMD calculations that are scaled up by 20\%. The proton yield has been corrected for weak decay feed-down from hyperons~\cite{star2023production}. The $\Lambda/p$ ratio increases from target rapidity to mid-rapidity, and also increases from peripheral to central collisions. Calculations from the UrQMD model reproduce the centrality and rapidity distributions well but underestimate the overall ratios by 20\%. Since the model reproduced the $\Lambda$ yields (see Fig.~\ref{dndy}), this means that proton yields are overpredicted by the model calculations.

\begin{figure}[htbp]
\centering
\centerline{\includegraphics[width=0.6\textwidth]{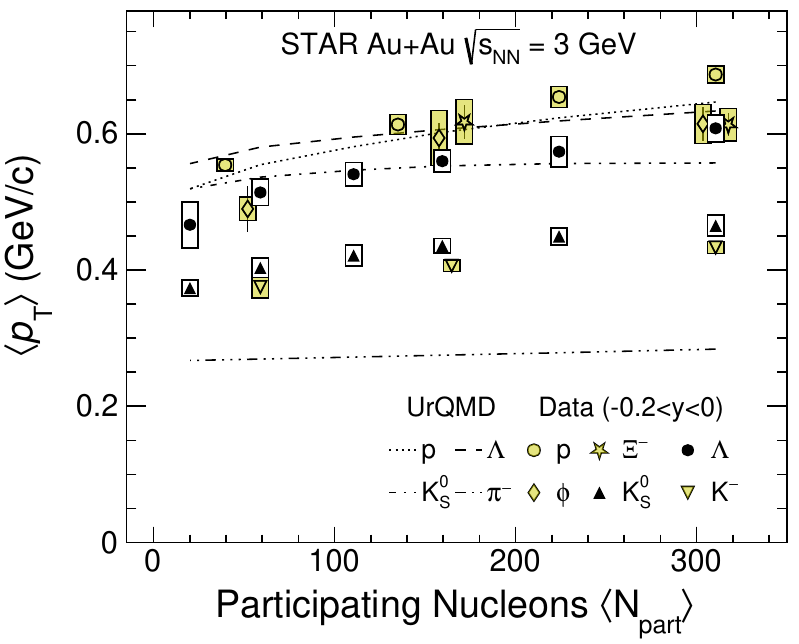}}
\caption{Mean transverse momentum \mpt for $\Xi^{-},~\Lambda,~\phi,~p,~K^0_S$, and $K^-$ at mid-rapidity ($-0.2<y_{\rm cm}<0$) as a function of \npart in Au+Au collisions at $\sqrt{s_{\rm NN}}$ = 3 GeV. The curves represent the calculations from UrQMD. Vertical lines represent statistical uncertainties, while boxes represent systematic uncertainties. $p$, $\phi$, $\Xi^{-}$, $K^-$ data are taken from Refs.~\cite{STAR:2021hyx,star2023production}.}
\label{meanpt}
\end{figure}
Figure~\ref{meanpt} shows \mpt as a function of the mean number of participating nucleons in a collision, \npart, for $K^-$~\cite{STAR:2021hyx}, $K^0_S$, $p$~\cite{star2023production}, $\Lambda$, $\phi$~\cite{STAR:2021hyx} and $\Xi^{-}$~\cite{STAR:2021hyx} at mid-rapidity ($-0.2<y_{\rm cm}<0$) in Au+Au collisions at $\sqrt{s_{\rm NN}}$ = 3 GeV. A gradual increase in \mpt with increasing \npart is observed for all particles which reflects stronger multiple scattering in central compared to peripheral collisions. The \mpt for protons is larger than for $\Lambda$s in central collisions even though the proton has a smaller mass than $\Lambda$. The effective temperature of protons in 0-10\% collisions at mid-rapidity, extracted via a fit using the Boltzmann function, is found to be $0.1629 \pm 0.0011$ GeV, which is larger than that of $\Lambda$s ($0.1469 \pm 0.0005$ GeV). Similar observations have been reported at similar collision energies~\cite{FOPI:2007usx, PhysRevLett.88.062301}. We note that $\Lambda$s are produced particles, while a fraction of protons arises directly from the incoming nucleons at such low energies. Comparing different strange hadrons, the data indicate $\mpt_{K^-}$ \apx $\mpt_{K^0_S}$ $<$ $\mpt_{\phi}$ \apx $\mpt_{\Lambda}$ \apx $\mpt_{\Xi^-}$, which approximately follows mass ordering. The ordering is consistent with collective radial flow caused by rescattering. The \mpt of $\Lambda$ is close to $\phi$ and $\Xi^-$ and show a deviation from the trend defined by $K^-$ and $p$. Similar trends have been observed at higher collision energies, as shown in Fig.~\ref{fig:NpartMass}, which may be due to strange hadrons having a smaller scattering cross section compared to ordinary hadrons in the later hadronic stage of the collisions~\cite{abelev2009measurements}. The increase in \mpt from peripheral to central collisions, and the mass ordering of \mpt for strange hadrons suggest the importance of hadronic rescatterings at this energy, and may be interpreted as evidence of collectivity. Note that by collectivity we mean the combined motion of observed hadrons that results, for example, in observables v1, v2~\cite{STAR:2021yiu} and the hadron mass and 
collision centrality dependence of mean transverse momentum \mpt, see in Fig.~\ref{meanpt}. The hadronic transport model UrQMD approximately reproduces the trends in mass and centrality dependences of \mpt for all particles. The \mpt of $K^0_S$ is overpredicted by UrQMD by $\sim 20\%$. In \snn = 2.4 GeV Au+Au collisions, $K^0_S$ is produced below the threshold and its mean transverse momentum is also overpredicted by UrQMD~\cite{HADES:2018noy}. Further studies are called for in order to understand the source of the discrepancy and the underlying mechanism for strangeness production in such collisions.

\begin{figure}[htbp]
\centering
\centerline{\includegraphics[width=0.55\textwidth]{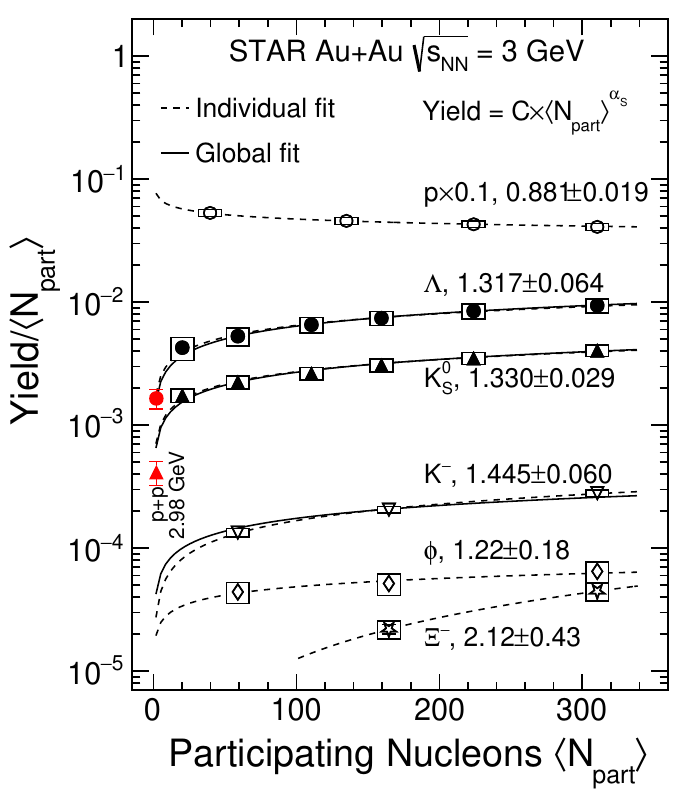}}
\caption{Hadron 4$\pi$ yields per mean number of participants Yield/\npart as a function of \npart. Vertical lines represent statistical uncertainties, while boxes represent systematic uncertainties. Power-law fits are shown as dashed lines for each particle. The result of a simultaneous fit to $S=1$ hadron yields ($K^-,~K^0_S,~\Lambda$) gives $\alpha_S = 1.348\pm0.028$ with $\rm \chi^2/NDF$ = 7/11, which is shown as solid lines. Data ($p$, $K^{-}$, $\phi$ and $\Xi^{-}$) are taken from~\cite{STAR:2021hyx,star2023production}. For comparison, $\Lambda$ and $K^0_S$ yields in $p$+$p$ collisions at $\sqrt{s_{\rm{NN}}}=2.98$ GeV ~\cite{kolesnikov2020new} are also shown as red symbols.}
\label{mult_vs_npart}
\end{figure}

With the wide rapidity coverages for all measured hadrons in the STAR FXT setup, 4$\pi$ yields can be readily estimated by fitting $dN/dy$ distribution. The 4$\pi$ yields of $K^-$~\cite{STAR:2021hyx}, $K^0_S$, $\Lambda$ as well as $p$~\cite{star2023production}, $\phi$~\cite{STAR:2021hyx} and $\Xi^{-}$~\cite{STAR:2021hyx}, normalized with the mean number of participants Yield/\npart, are shown in Fig.~\ref{mult_vs_npart} as a function of \npart. To quantify the centrality dependence, yields of  $K^-,~K^0_S,~\Lambda$ are fitted with a function $f=C \langle \rm N_{part} \rangle^{\alpha_S}$, where $C$ is a constant, and $\alpha_S$ is a power-law scaling parameter~\cite{HADES:2018noy}. The $\alpha_S$ parameters for $S=1$ hadrons $(\Lambda,K_S^{0}, K^{-})$ are consistent within uncertainties. A common fit to these three hadrons leads to the result $\alpha_S = 1.348\pm0.028$. A similar result, $1.383\pm0.021$, is obtained if we only use mid-rapidity yields ($-0.5<y_{\rm cm}<0$) for fitting. This common scaling between $S=1$ hadrons suggests a similar production mechanism for these particles. 

For comparison, the yields of $p$, $\phi$ and $\Xi^{-}$ are also shown in Fig.~\ref{mult_vs_npart}. The centrality dependence of $\phi$ is consistent with that of the $S=1$ hadrons, while $p$ and $\Xi^{-}$ deviate from the scaling, indicating a different production mechanism. At this energy $\sqrt{s_{\rm{NN}}}=3$ GeV, most protons are not produced but are remnants from the incoming nuclei, which explains the smaller $\alpha_S$. Meanwhile, the $\alpha_S$ for $\Xi^{-}$ is larger compared to that for $S=1$  hadrons by $\sim$ 1.8$\sigma$. Similar results were reported for $\sqrt{s_{\rm{NN}}}=3.45$ GeV Au+Au collisions~\cite{E895:2003qcm} and this difference is also seen in UrQMD. The multi-strange baryon $\Xi^-$ has a $NN$-production threshold of 3.25 GeV. Its value of $\alpha_S \sim 2$ is larger than for other strange hadrons which may be reflecting the effect of sub-threshold production~\cite{Song:2020clw}. The yields of $K^0_S$ and $\Lambda$ are also measured in $\sqrt{s_{\rm{NN}}}=2.98$ GeV $p$+$p$ collisions~\cite{kolesnikov2020new} and shown as red symbols in Fig.~\ref{mult_vs_npart}. These yields are consistent with \npart scaling within 3$\sigma$. In general, $\alpha_S$ is larger than 1 for strange hadrons, indicating an increase of normalized yields from peripheral to central collisions. This behavior is also observed in UrQMD calculations, which yield $\alpha_S = 1.45$. This increase may be attributed to hadronic rescatterings involving baryonic resonances that dominantly contribute to the production of strange particles near threshold~\cite{Miskowiec:1994vj, FOPI:2007usx, KaoS:1997, E-802:1998}.

\begin{figure}[htb]
\centering
\centerline{\includegraphics[width=0.6\textwidth]{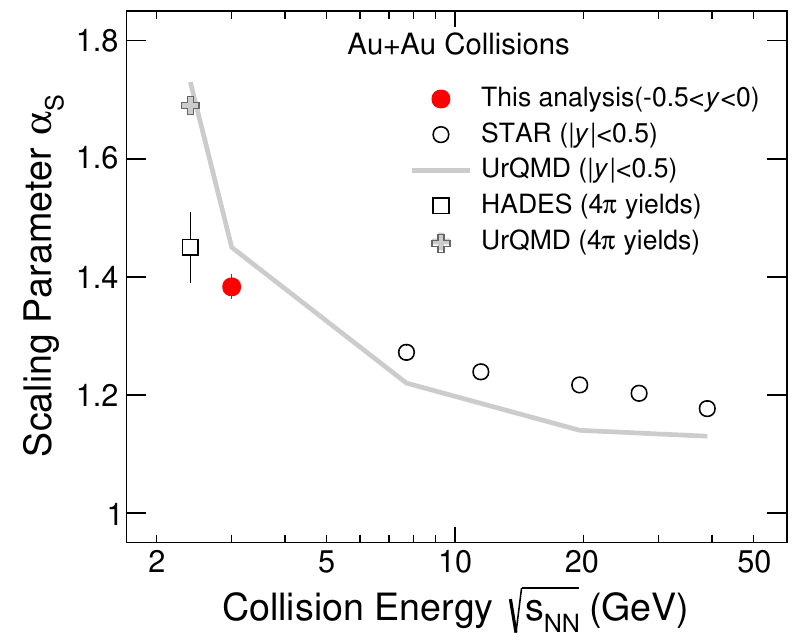}}
\caption{Energy dependence of the strange hadron production scaling parameter $\alpha_S$. STAR data for mid-rapidity yields are shown as open circles ($|y_{\rm cm}|\leq 0.5$)~\cite{STAR:2019bjj} and filled circles ($-0.5\leq y_{\rm cm} \leq 0$). The open square is the 4$\pi$ strange hadron yield from HADES~\cite{HADES:2018noy}.  }
\label{fig:scaling}
\end{figure}

To further interpret the data, in Fig.~\ref{fig:scaling} we investigate the energy dependence of the common power-law scaling of strange hadron yields. Strange particle yields $(\Lambda,K_S^{0})$ in $\sqrt{s_{\rm{NN}}}=7.7,~11.5,~19.6,~27$ and $39$ GeV collisions~\cite{STAR:2019bjj} are used to extract $\alpha_S$ at these energies and compared to the present result at $\sqrt{s_{\rm{NN}}}=$ 3 GeV and to the $\sqrt{s_{\rm{NN}}}=$ 2.4 GeV results from HADES~\cite{HADES:2018noy}. As shown in Fig.~\ref{fig:scaling}, the power-law scaling parameter $\alpha_S$ decreases with collision energy. Above $\sqrt{s_{\rm{NN}}}= 10$ GeV, the rate of decrease becomes much slower for most of the produced hadrons and appears to saturate at high beam energies~\cite{STAR:2019bjj} ($\alpha_S = 1.1 \pm 0.03$ in $\sqrt{s_{\rm{NN}}}=$ 200 GeV Au+Au collisions~\cite{STAR:2006egk} and $1.15 \pm 0.02$ in $\sqrt{s_{\rm{NN}}}=$ 2.76 TeV Pb+Pb collisions~\cite{Belikov:2011xk}). Calculations of mid-rapidity strange hadron production from the UrQMD model, indicated by the solid gray line in Fig.~\ref{fig:scaling}, reproduce the decreasing slope in the energy dependence, although the trend is steeper in the model compared to the data. At low collision energies, the increase of $\alpha_S$ is partly due to sub-threshold production which may be sensitive to the EoS~\cite{HADES:2018noy}, while at higher energies, thermal production of strangeness in the QGP is not included in UrQMD and may explain the differences between data and calculations.

\begin{figure}[htbp]
\centering
\centerline{\includegraphics[width=0.6\textwidth]{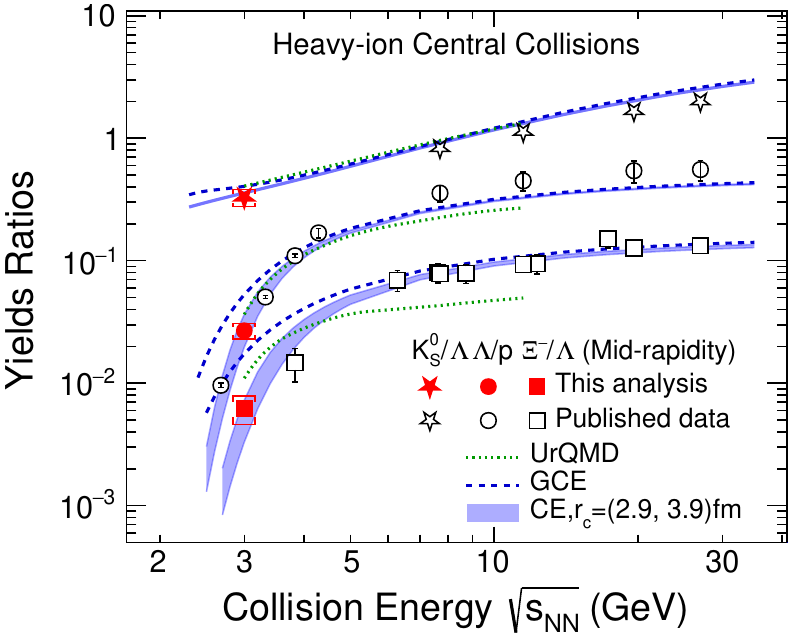}}
\caption{Mid-rapidity yield ratios of $K^0_S/\Lambda$ (stars), $\Lambda/p$ (circles) and $\Xi^-/\Lambda$ (squares) as a function of collision energy $\sqrt{s_{\rm{NN}}}$. The new results for top 10\% central Au+Au collisions at $\sqrt{s_{\rm{NN}}}=3$ GeV are shown as red filled symbols, while empty markers in black are used for data from various other energies~\cite{E895:2001zms,E895:2001yfr,E895:2003qcm,NA49:2008ysv,STAR:2019bjj}. Vertical lines and boxes represent statistical and systematic uncertainties, respectively. Blue hatched bands are calculations from THERMUS~\cite{wheaton2009thermus} with canonical ensemble using a strangeness correlation radius $r_{\rm c}$ ranging from 2.9 to 3.9 fm. Blue dashed lines are results from the same model but with the grand canonical ensemble. Green dotted lines show calculations from UrQMD~\cite{Bleicher:1999xi} for central Au+Au collisions.} \label{ratio_vs_e}
\end{figure}

The $\Lambda$ yield as a function of energy is non-monotonic~\cite{STAR:2019bjj}, with a local minimum near $\sqrt{s_{\rm{NN}}}=39$ GeV. This behavior is believed to originate from an interplay between pair production, which strongly increases with increasing collision energy, and the associated production of $\Lambda$ in nucleon-nucleon scatterings, which strongly increases with increasing net baryon density. The proton $dN/dy$ also shows a similar minimum at 39 GeV~\cite{STAR:2013gus}, which may suggest that $\Lambda$s are predominantly created via associated production from protons and neutrons in this energy range. To cancel baryon density effects, we study the ratios $\Lambda/p$ and $\Xi^-/\Lambda$. The energy dependence of the mid-rapidity yield ratios of $\Lambda/p$ (circles) and $\Xi^-/\Lambda$ (squares) is shown in Fig.~\ref{ratio_vs_e}, including the midrapidity data in central Au+Au or Pb+Pb data from the AGS, SPS and RHIC BES at higher energies~\cite{E895:2001zms,E895:2001yfr,E895:2003qcm,NA49:2008ysv,STAR:2019bjj}. The new results from top 10\% central Au+Au collisions at $\sqrt{s_{\rm{NN}}}=3$ GeV are shown as red filled symbols. The $\Lambda/p$ and $\Xi^-/\Lambda$ ratios decrease rapidly with decreasing collision energy. This rapid decrease is primarily due to two effects. The first is the elementary $NN$ production threshold. The production rates of strange hadrons $\Lambda$ and $\Xi^-$ will drop rapidly when the beam energy falls below the elementary $NN$ production threshold, while the proton $dN/dy$ will increase as the collision energy decreases due to baryon stopping at lower collision energies. The second effect is canonical suppression which leads to an additional suppression for strange hadrons at lower incident energies~\cite{STAR:2021hyx}. The $K^0_S/\Lambda$ ratio increases monotonically with collision energy, suggestive of a transition from baryon-dominated matter at lower energies to a meson-dominated matter at higher energies.

As one can see in the figure, UrQMD cannot quantitatively reproduce the energy dependence of the measured ratios $\Lambda/p$ and $\Xi^-/\Lambda$. This may indicate that important mechanisms for strange hadron production are still missing in the model calculations, e.g. feed-down from high-mass baryon resonances~\cite{Steinheimer_2015_UrQMD}.

Statistical thermal model calculations, which assume thermal and chemical equilibrium at freeze-out, have been widely used to characterize the properties of the produced medium in heavy-ion collisions. The strange hadron ratios from thermal model THERMUS with Canonical Ensemble (CE), using a strangeness correlation radius $r_c \sim$ $2.9-3.9$\,fm, are obtained and depicted by blue hatched bands in Fig.~\ref{ratio_vs_e}. Results from the same model calculations with Grand Canonical Ensemble (GCE), blue dashed lines, are also shown for comparison. In the present analysis, the freeze-out parameters, including the chemical freeze-out temperature $T_{\rm ch}$ and the  baryon chemical potential $\mu_{\rm B}$ are taken from Ref.~\cite{Vovchenko:2015idt}, with $\mu_{\rm S}$ constrained by requiring zero net strangeness. Both CE and GCE can describe the $K^0_S/\Lambda$ data and they are very similar in the whole energy range. For $\Lambda/p$ and $\Xi^-/\Lambda$, both CE and GCE results converge and are consistent with experimental observations at higher collision energies, while only the calculations with CE could reproduce the ratios at lower collision energies. The failure of GCE to describe strangeness production at lower collision energy suggests that at lower collision energies where the multiplicity is low, there is a local/exact conservation of charge, baryon and strangeness quantum numbers. 

It is worth noting that the strangeness correlation radii of $r_c \sim$ $2.9-3.9$\,fm are used in order to simultaneously describe the measured $\Lambda/p$, $\Xi^-/\Lambda$ and $K^0_S/\Lambda$ data. The Thermal-FIST model~\cite{thermalfist} with CE can also reproduce the data, albeit with a slightly different $r_c \sim$ $3.2-3.9$\,fm. However, as discussed in Ref.~\cite{STAR:2021hyx}, different radii of  $r_c\sim2.7$\,fm ($r_c\sim4.2$\,fm) are needed to reproduce the ratio of $\phi/K^-$($\phi/\Xi^-$)~\cite{STAR:2021hyx} from the same collisions. Further investigation, including the precise determination of the chemical freeze-out parameters $T_{\rm ch}$, $\mu_{\rm B}$, and strangeness suppression factor $\gamma_s$~\cite{Tawfik:2014dha} through a global fit to all measured particle yields from $\sqrt{s_{\rm NN}}$ = 3\,GeV Au+Au collisions, should be conducted in the future. 

\section{Summary}
\label{sec:summary}
In summary, we report multi-differential yield measurements of $\Lambda$ and $K^0_S$ in Au+Au collisions at $\sqrt{s_{\rm NN}}$ = 3 GeV with the STAR experiment at RHIC. Clear centrality dependence of the average transverse momentum of strange particles ($K^-$, $K^0_S$, $\phi$, $\Lambda$, $\Xi^{-}$) has been observed, providing evidence of hadronic collectivity due to hadronic rescatterings in such collisions. The 4$\pi$ yields of strange hadrons follow a common centrality dependence within experimental uncertainties except for $\Xi^{-}$. This discrepancy could be due to the sub-threshold production for $\Xi^{-}$ at this collision energy. The extracted strange hadron scaling parameter $\alpha_S$ is consistent with a monotonically decreasing energy dependence within $2.4 \leq \sqrt{s_{\rm NN}}\leq 40$\,GeV. The hadronic transport model UrQMD qualitatively describes the decreasing trend in $\alpha_{S}$, while also describing the trends in the centrality and rapidity dependence observed in the $\sqrt{s_{\rm NN}}$ = 3\,GeV data.
In contrast to the situation in high energy collisions where the thermal model with GCE  describes particle production well, at 3\,GeV, CE with $r_{\rm c}$ from 2.9 to 3.9 fm simultaneously describes the mid-rapidity ratios $K^0_S/\Lambda$, $\Lambda/p$, and $\Xi^-/\Lambda$ in central collisions. Similar features in the ratios $\phi/K^-$ and $\phi/\Xi^-$ have also been observed~\cite{STAR:2021hyx}. The change from GCE to CE, reflected in the strange particle ratios, in addition to the fact that a hadronic transport model reproduces the results, imply the dominance of hadronic interactions in the EoS of the medium created in $\sqrt{s_{\rm NN}}$ = 3\,GeV Au+Au collisions.

\appendix
\section{Supplementary material}
\label{sec:suppl}

\begin{figure}[htbp]
\centering
\centerline{\includegraphics[width=0.95\textwidth]{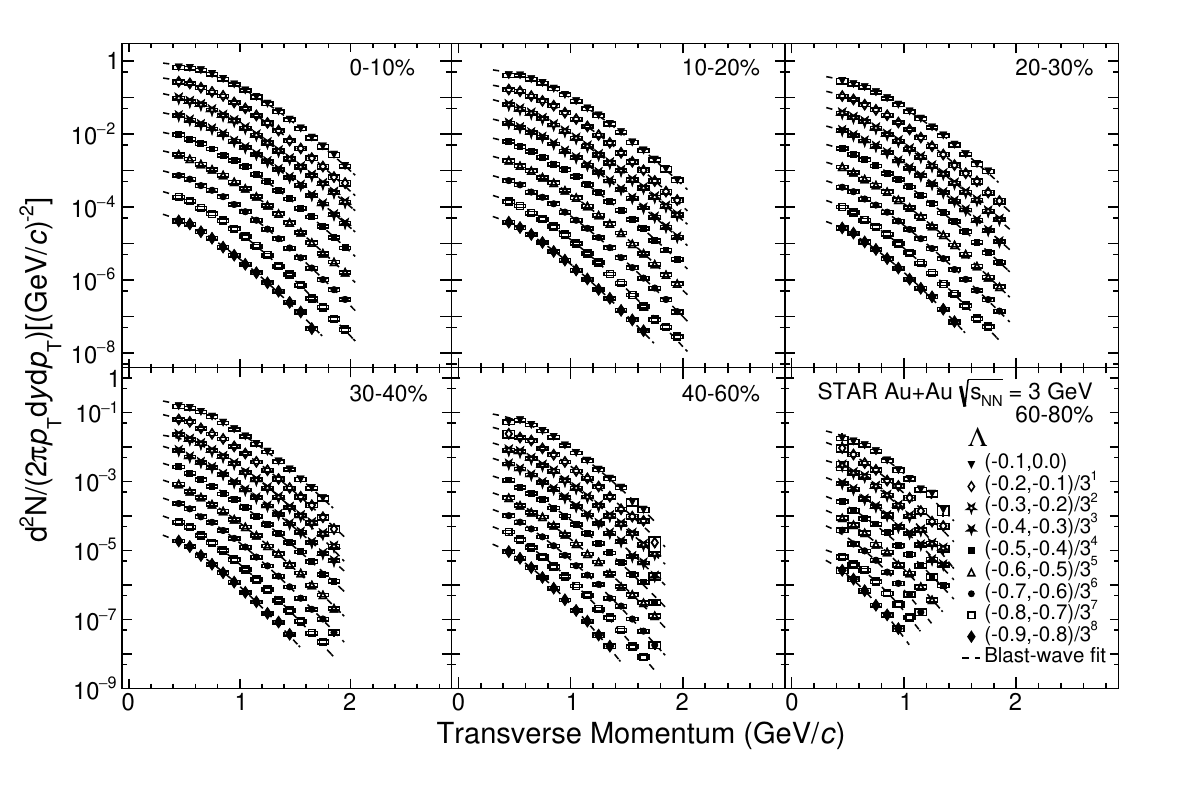}}
\caption{The transverse-momentum spectra of $\Lambda$ for different rapidities and centralities in Au+Au collisions at $\sqrt{s_{\rm NN}}$ = 3 GeV. The spectra have been corrected for the feed-down from $\Xi^-$ and $\Xi^0$ decays. The data points are scaled by factors of $1/3$ from mid- to forward rapidities as indicated in the legend. The vertical lines and boxes represent the statistical and systematic uncertainties, respectively. The dashed curves represent fits to the data using the blast-wave model.}
\label{ld_pt_all}
\end{figure}

\begin{figure}[htbp]
\centering
\centerline{\includegraphics[width=0.95\textwidth]{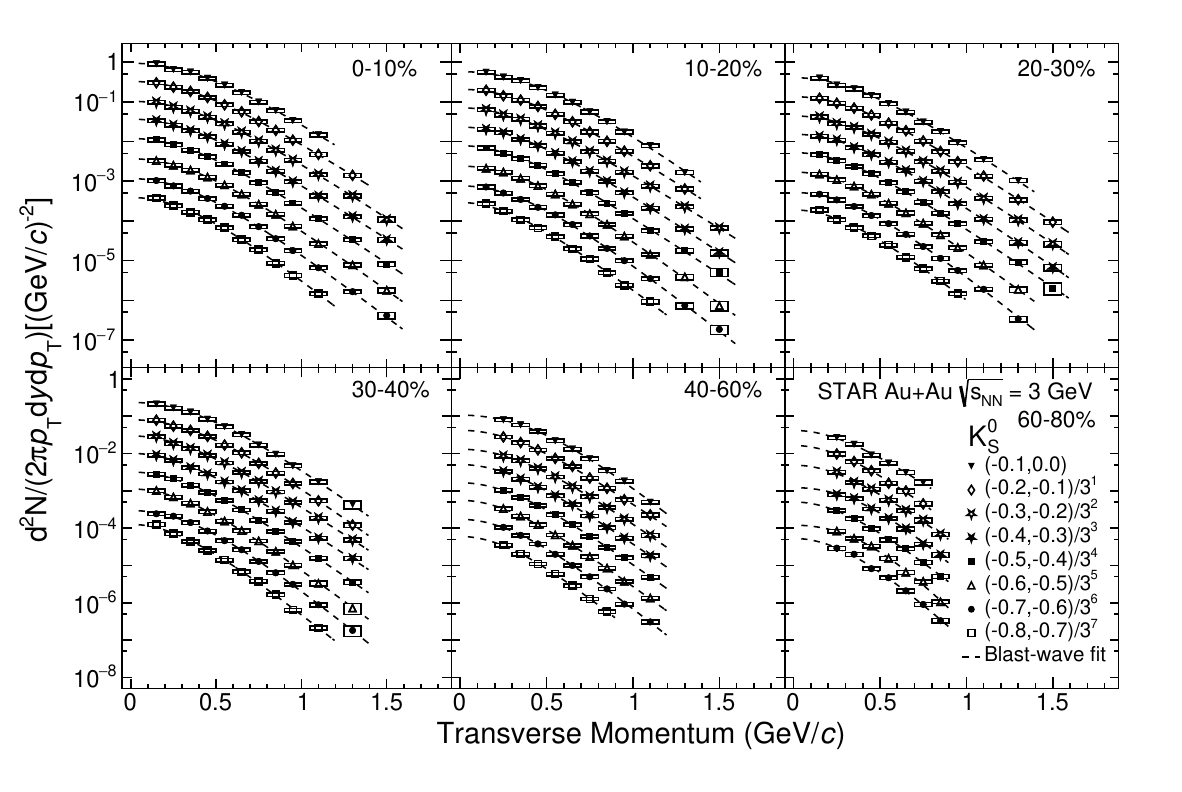}}
\caption{The transverse-momentum spectra of $K^0_S$ for different rapidities and centralities in Au+Au collisions at $\sqrt{s_{\rm NN}}$ = 3 GeV. The data points are scaled by factors of $1/3$ from mid- to forward rapidities as indicated in the legend. The vertical lines and boxes represent the statistical and systematic uncertainties, respectively. The dashed curves represent fits to the data using the blast-wave model.}
\label{ks_pt_all}
\end{figure}

\begin{table}[htbp]
\centering
\caption{Integral yield ($dN/dy$) of $\Lambda$ at different centralities in Au+Au collisions at $\sqrt{s_{\rm NN}}$ = 3\,GeV. The first uncertainty is statistical, and the second is systematic.}
\begin{tabular}{c|c|c|c}
\hline
Rapidity & 0-10\% & 10-20\% & 20-30\%  \\
\hline
y[-0.9, -0.8] & $0.561 \pm 0.007 \pm 0.083$ & $0.468 \pm 0.006 \pm 0.081$ & $0.320 \pm 0.004 \pm 0.061$ \\
y[-0.8, -0.7] & $0.815 \pm 0.008 \pm 0.125$ & $0.603 \pm 0.006 \pm 0.095$ & $0.413 \pm 0.004 \pm 0.062$ \\
y[-0.7, -0.6] & $1.093 \pm 0.009 \pm 0.161$ & $0.757 \pm 0.007 \pm 0.109$ & $0.499 \pm 0.004 \pm 0.070$ \\
y[-0.6, -0.5] & $1.359 \pm 0.011 \pm 0.192$ & $0.890 \pm 0.007 \pm 0.126$ & $0.572 \pm 0.005 \pm 0.078$ \\
y[-0.5, -0.4] & $1.586 \pm 0.013 \pm 0.224$ & $0.996 \pm 0.008 \pm 0.138$ & $0.615 \pm 0.005 \pm 0.082$ \\
y[-0.4, -0.3] & $1.836 \pm 0.015 \pm 0.253$ & $1.119 \pm 0.010 \pm 0.151$ & $0.678 \pm 0.006 \pm 0.089$ \\
y[-0.3, -0.2] & $2.061 \pm 0.019 \pm 0.248$ & $1.251 \pm 0.012 \pm 0.147$ & $0.724 \pm 0.007 \pm 0.089$ \\
y[-0.2, -0.1] & $2.262 \pm 0.015 \pm 0.244$ & $1.355 \pm 0.009 \pm 0.164$ & $0.772 \pm 0.005 \pm 0.084$ \\
y[-0.1,  0.0] & $2.258 \pm 0.017 \pm 0.267$ & $1.329 \pm 0.010 \pm 0.173$ & $0.752 \pm 0.006 \pm 0.088$ \\
\hline
Rapidity & 30-40\% & 40-60\% & 60-80\% \\
\hline
y[-0.9, -0.8] & $0.216 \pm 0.003 \pm 0.041$ & $0.107 \pm 0.002 \pm 0.024$ & $0.031 \pm 0.001 \pm 0.009$ \\
y[-0.8, -0.7] & $0.268 \pm 0.003 \pm 0.045$ & $0.122 \pm 0.002 \pm 0.024$ & $0.027 \pm 0.001 \pm 0.006$ \\
y[-0.7, -0.6] & $0.309 \pm 0.003 \pm 0.047$ & $0.133 \pm 0.002 \pm 0.025$ & $0.040 \pm 0.001 \pm 0.009$ \\
y[-0.6, -0.5] & $0.352 \pm 0.003 \pm 0.047$ & $0.142 \pm 0.002 \pm 0.023$ & $0.037 \pm 0.001 \pm 0.008$ \\
y[-0.5, -0.4] & $0.376 \pm 0.004 \pm 0.050$ & $0.153 \pm 0.002 \pm 0.023$ & $0.035 \pm 0.001 \pm 0.007$ \\
y[-0.4, -0.3] & $0.399 \pm 0.004 \pm 0.054$ & $0.167 \pm 0.002 \pm 0.025$ & $0.041 \pm 0.001 \pm 0.007$ \\
y[-0.3, -0.2] & $0.422 \pm 0.005 \pm 0.057$ & $0.165 \pm 0.002 \pm 0.024$ & $0.052 \pm 0.002 \pm 0.010$ \\
y[-0.2, -0.1] & $0.436 \pm 0.003 \pm 0.053$ & $0.172 \pm 0.002 \pm 0.022$ & $0.047 \pm 0.001 \pm 0.010$ \\
y[-0.1,  0.0] & $0.425 \pm 0.004 \pm 0.048$ & $0.177 \pm 0.002 \pm 0.022$ & $0.045 \pm 0.001 \pm 0.009$ \\
\hline
\end{tabular}
\end{table}

\begin{table*}[htb]
\centering
\caption{Integral yield ($dN/dy$) of $K^{0}_{S}$ at different centralities in Au+Au collisions at $\sqrt{s_{\rm NN}}$ = 3\,GeV. The first uncertainty is statistical, and the second is systematic.}
\begin{tabular}{c|c|c|c}
\hline
Rapidity & 0-10\% & 10-20\% & 20-30\% \\
\hline
y[-0.8, -0.7] & $0.456 \pm 0.011 \pm 0.047$ & $0.302 \pm 0.008 \pm 0.031$ & $0.190 \pm 0.006 \pm 0.020$ \\
y[-0.7, -0.6] & $0.509 \pm 0.007 \pm 0.052$ & $0.317 \pm 0.005 \pm 0.032$ & $0.201 \pm 0.004 \pm 0.021$ \\
y[-0.6, -0.5] & $0.578 \pm 0.007 \pm 0.059$ & $0.355 \pm 0.004 \pm 0.036$ & $0.223 \pm 0.003 \pm 0.023$ \\
y[-0.5, -0.4] & $0.645 \pm 0.006 \pm 0.065$ & $0.387 \pm 0.004 \pm 0.039$ & $0.240 \pm 0.003 \pm 0.024$ \\
y[-0.4, -0.3] & $0.690 \pm 0.006 \pm 0.070$ & $0.408 \pm 0.004 \pm 0.042$ & $0.251 \pm 0.003 \pm 0.026$ \\
y[-0.3, -0.2] & $0.709 \pm 0.006 \pm 0.073$ & $0.425 \pm 0.004 \pm 0.044$ & $0.256 \pm 0.003 \pm 0.026$ \\
y[-0.2, -0.1] & $0.740 \pm 0.006 \pm 0.077$ & $0.436 \pm 0.004 \pm 0.045$ & $0.267 \pm 0.003 \pm 0.028$ \\
y[-0.1,  0.0] & $0.742 \pm 0.007 \pm 0.079$ & $0.437 \pm 0.005 \pm 0.047$ & $0.264 \pm 0.003 \pm 0.030$ \\
\hline
Rapidity & 30-40\% & 40-60\% & 60-80\% \\
\hline
y[-0.8, -0.7] & $0.119 \pm 0.005 \pm 0.012$ & $0.052 \pm 0.002 \pm 0.005$ & $0.015 \pm 0.001 \pm 0.002$ \\
y[-0.7, -0.6] & $0.116 \pm 0.003 \pm 0.012$ & $0.055 \pm 0.001 \pm 0.006$ & $0.015 \pm 0.001 \pm 0.002$ \\
y[-0.6, -0.5] & $0.134 \pm 0.002 \pm 0.014$ & $0.060 \pm 0.001 \pm 0.006$ & $0.017 \pm 0.001 \pm 0.002$ \\
y[-0.5, -0.4] & $0.142 \pm 0.002 \pm 0.014$ & $0.064 \pm 0.001 \pm 0.007$ & $0.017 \pm 0.001 \pm 0.002$ \\
y[-0.4, -0.3] & $0.149 \pm 0.002 \pm 0.015$ & $0.068 \pm 0.001 \pm 0.007$ & $0.019 \pm 0.001 \pm 0.002$ \\
y[-0.3, -0.2] & $0.155 \pm 0.002 \pm 0.016$ & $0.066 \pm 0.001 \pm 0.007$ & $0.021 \pm 0.001 \pm 0.002$ \\
y[-0.2, -0.1] & $0.156 \pm 0.002 \pm 0.016$ & $0.071 \pm 0.001 \pm 0.008$ & $0.020 \pm 0.001 \pm 0.002$ \\
y[-0.1,  0.0] & $0.155 \pm 0.002 \pm 0.018$ & $0.070 \pm 0.001 \pm 0.008$ & \\
\hline
\end{tabular}
\end{table*}

\begin{table}[htbp]
\centering
\caption{Mean transverse momentum \mpt of $\Lambda$ and $K^{0}_{S}$ at different centralities in Au+Au collisions at $\sqrt{s_{\rm NN}}$ = 3\,GeV. The first uncertainty is statistical, and the second is systematic.}
\begin{tabular}{c|c|c|c}
\hline
\mpt ($\rm{GeV}/c$) & 0-10\% & $10-20\%$ & $20-30\%$ \\
\hline
$\Lambda$   & $0.608 \pm 0.0015 \pm 0.020$ & $0.574 \pm 0.0015 \pm 0.022$ & $0.560 \pm 0.0016 \pm 0.016$ \\
$K^{0}_{S}$ & $0.465 \pm 0.001 \pm 0.017$ & $0.449 \pm 0.001 \pm 0.012$ & $0.436 \pm 0.001 \pm 0.012$ \\
\hline
\mpt ($\rm{GeV}/c$) & $30-40\%$ & $40-60\%$ & $60-80\%$ \\
\hline
$\Lambda$   & $0.541 \pm 0.0018 \pm 0.017$ & $0.514 \pm 0.0027 \pm 0.019$ & $0.466 \pm 0.006 \pm 0.034$ \\
$K^{0}_{S}$ & $0.422 \pm 0.001 \pm 0.016$ & $0.403 \pm 0.002 \pm 0.016$ & $0.369 \pm 0.004 \pm 0.011$ \\
\hline
\end{tabular}
\end{table}

\begin{figure}[htbp]
\centering
\centerline{\includegraphics[width=0.6\textwidth]{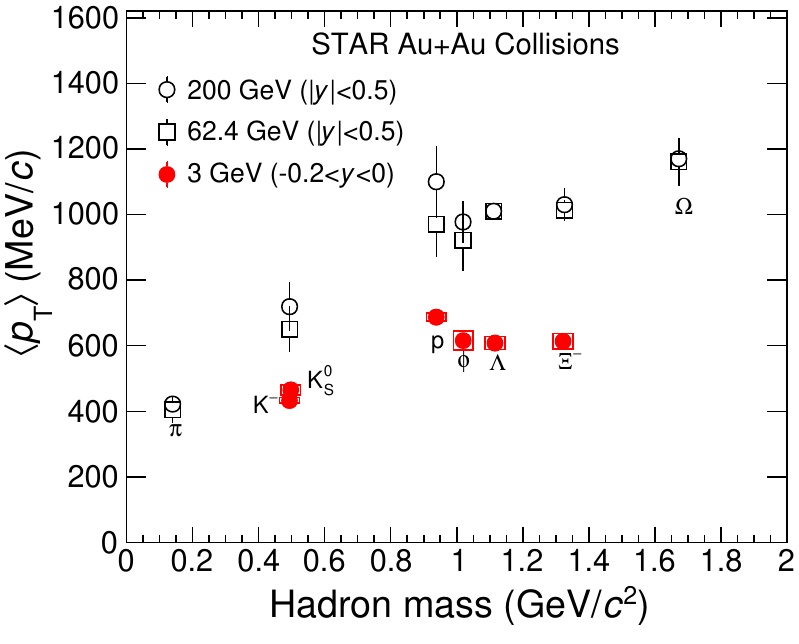}}
\caption{Hadron mass dependence of mean transverse momentum \mpt at mid-rapidity in central Au+Au collisions at $\sqrt{s_{\rm{NN}}}=3, 62.4$  and 200 GeV~\cite{abelev2009measurements}. The new results at \snn = 3 GeV are shown as red filled symbols. Vertical lines and boxes represent statistical and systematic uncertainties, respectively.}
\label{fig:NpartMass}
\end{figure}

\newpage
\acknowledgments
We thank the RHIC Operations Group and RCF at BNL, the NERSC Center at LBNL, and the Open Science Grid consortium for providing resources and support.  This work was supported in part by the Office of Nuclear Physics within the U.S. DOE Office of Science, the U.S. National Science Foundation, National Natural Science Foundation of China, Chinese Academy of Science, the Ministry of Science and Technology of China and the Chinese Ministry of Education, the Higher Education Sprout Project by Ministry of Education at NCKU, the National Research Foundation of Korea, Czech Science Foundation and Ministry of Education, Youth and Sports of the Czech Republic, Hungarian National Research, Development and Innovation Office, New National Excellency Programme of the Hungarian Ministry of Human Capacities, Department of Atomic Energy and Department of Science and Technology of the Government of India, the National Science Centre and WUT ID-UB of Poland, the Ministry of Science, Education and Sports of the Republic of Croatia, German Bundesministerium f\"ur Bildung, Wissenschaft, Forschung and Technologie (BMBF), Helmholtz Association, Ministry of Education, Culture, Sports, Science, and Technology (MEXT), Japan Society for the Promotion of Science (JSPS) and Agencia Nacional de Investigaci\'on y Desarrollo (ANID) of Chile.

\newpage
\bibliographystyle{JHEP}
\bibliography{biblio.bib}







\end{document}